\documentclass[10pt, twocolumn, longbibliography,aps,groupedaddress, superscriptaddress,notitlepage]{revtex4-2}
\usepackage[colorlinks=true, linkcolor=blue, citecolor=blue, urlcolor=blue ]{hyperref}
\usepackage{graphicx} 		 
\usepackage[none]{hyphenat}		
\usepackage{braket}

\begin{document}
\title{Ultra-sensitive measurement of transverse displacements with structured light}

\author{Raouf Barboza}
\affiliation{Dipartimento di Fisica, Universit\`a di Napoli Federico II, Complesso Universitario di Monte S. Angelo, Via Cintia, 80126 Napoli, Italy}

\author{Amin Babazadeh}
\affiliation{Dipartimento di Fisica, Universit\`a di Napoli Federico II, Complesso Universitario di Monte S. Angelo, Via Cintia, 80126 Napoli, Italy}

\author{Lorenzo Marrucci}
\affiliation{Dipartimento di Fisica, Universit\`a di Napoli Federico II, Complesso Universitario di Monte S. Angelo, Via Cintia, 80126 Napoli, Italy}

\author{Filippo Cardano}
\email[]{filippo.cardano2@unina.it}
\affiliation{Dipartimento di Fisica, Universit\`a di Napoli Federico II, Complesso Universitario di Monte S. Angelo, Via Cintia, 80126 Napoli, Italy}

\author{Corrado de Lisio}
\affiliation{Dipartimento di Fisica, Universit\`a di Napoli Federico II, Complesso Universitario di Monte S. Angelo, Via Cintia, 80126 Napoli, Italy}
\affiliation{CNR-SPIN U.O.S. di Napoli, Via Cintia, 80126 Napoli, Italy}

\author{Vincenzo D'Ambrosio}
\email[]{vincenzo.dambrosio@unina.it}
\affiliation{Dipartimento di Fisica, Universit\`a di Napoli Federico II, Complesso Universitario di Monte S. Angelo, Via Cintia, 80126 Napoli, Italy}

\begin{abstract}
Accurately measuring mechanical displacements is essential for a vast portion of current technologies. Several optical techniques accomplish this task, allowing for non-contact sensing even below the diffraction limit. Here we introduce an optical encoding technique, dubbed ``linear photonic gears'', that enables ultra-sensitive measurements of transverse displacements by mapping these into polarization rotations of a laser beam. In ordinary ambient conditions, we measure the relative shift between two objects with a resolution of 400~pm. We argue that a resolution of 50~pm should be achievable with existing state-of-the-art technologies. Our single-optical-path scheme is intrinsically stable and it could be implemented as a compact sensor, using integrated optics. We anticipate it may have a strong impact on both research and industry.
\end{abstract}

\maketitle

Reading-out and tracking precisely the position of a system is of key relevance in fields as different as microscopy, mechanical engineering, quantum physics, material science, semiconductor industry or general relativity \cite{grav,afm,Bier10,Knob03,Miao20}. To this end, light has emerged as an invaluable tool, as it allows for fast, non-invasive and accurate sensing \cite{deGroot2019}.  
In photonic systems, displacements can be regarded as either parallel or transverse to the main propagation direction of the optical beam. While in the first case triangulation measurements or interferometric setups \cite{Berk12} can be used, the measurement of transverse displacements (TD) typically relies on the detection of differential current signals from photodiodes \cite{differential}. This provides a practical but limited solution in terms of sensitivity and resolution. For improved performance other techniques are available, which exploit for instance grating interferometry \cite{GI}, diffraction based overlay \cite{dbo} or fluorophore imaging \cite{fluo}. However, while yielding subnanometric resolutions, these methods are often based on complex apparatus and are therefore limited in terms of versatility.

Structured light \cite{roadmap}, that is an optical field presenting a spatially-varying distribution of amplitude, phase and/or polarization, emerged recently as a resource in this area \cite{Bag18,Tish18,Shan19,Yuan19,Bag20}. By exploiting structured illumination, TD can be indeed measured, for instance, via position dependent directional scattering from a nanoantenna \cite{Bag18,Shan19,Bag20} or via centroid tracking of the scattered field distribution \cite{Tish18}.
Moreover, by properly sculpting the phase profile of a light beam via a metasurface, an optical ruler exploiting super-oscillations achieves a resolution far below the diffraction limit \cite{Yuan19}.
Although these methods enable TD measurements with sub-nanometric resolutions, they all rely on high-magnification imaging systems and require one to match light wavelength to specific nano-antenna resonances or to post-process images via reconstruction algorithms. These factors impose limitations in terms of footprint, versatility, cost and speed, all relevant features of an ideal sensor, besides its sensitivity. Importantly, structured light can be a resource for enhanced sensing purposes even without high-magnification imaging, as for instance in the ``photonic gears'', in which  a bidirectional mapping between the polarization state and a properly tailored vectorial mode of a paraxial light beam enables a boost of the sensitivity in roll angle measurements \cite{gears13}. 

By combining a similar principle with a Moir\'e-like sensing scheme \cite{Moire,Moire2,Hane85} here we present and demonstrate a novel optical encoding method, the {\it linear photonic gears}, that enables an enhanced sensitivity in TD measurements with a compact, fast, stable and cost-effective setup. 
The linear gears concept is based on mapping the displacement into a rotation of the optical polarization thanks to a pair of Pancharatnam-Berry phase gratings. As such, the transverse displacement can be immediately retrieved by simply monitoring the optical power after a linear polarizer. The sensitivity of the method is controlled by the polarization rotation-rate per unit length in the direction of the TD.  
We experimentally test this principle with a minimal setup, in ordinary ambient conditions and at room temperature, and report nanometric TD measurements with a sub-nanometric resolution, mainly limited by the mechanical stability of our setup. In a more controlled, yet still realistic setting, we argue that the resolution could be reduced to the tens-of-picometer scale, comparable to the typical size of an atom.

\begin{figure*}[h!!t!]
\includegraphics[width=0.95\textwidth,keepaspectratio=true]{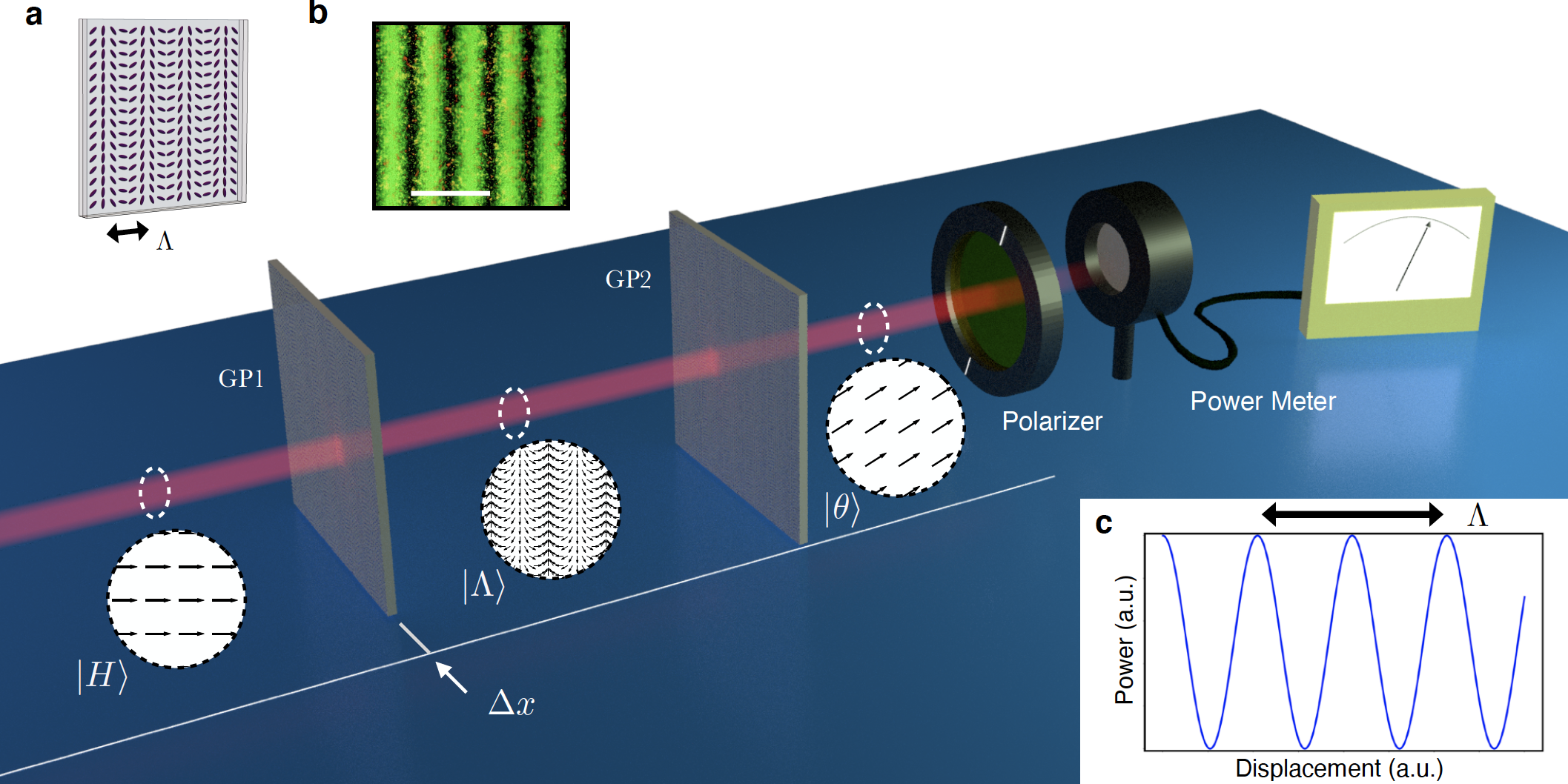}
\caption{Linear gears concept.  A laser beam, linearly polarized along the horizontal direction, is transformed by a $g$-plate (GP1) into a structured light beam $\ket{\Lambda}$. A second $g$-plate (GP2), identical to GP1, transforms back the polarization mode into a linearly polarized homogeneous one ($\ket{\theta}$), rotated by an angle $\Delta\theta$ with respect to the input polarization direction. Dashed circles represent the polarization state in three different positions along the laser beam corresponding to $\ket{H}$, $\ket{\Lambda}$ and  $\ket{\theta}$ respectively.  By simply monitoring the optical power after a linear polarizer, it is possible to keep track of the transverse displacement $\Delta x$.   a) $g$-plate geometry,  blue lines represent the orientation of the liquid-crystal molecular director inside the cell. The spatial period is set by the parameter $\Lambda$. b) Microscope picture of a $g$-plate with $\Lambda = 50$ $\mu m$ placed between two linear polarizers. The scale bar corresponds to $50$ $\mu$m. c) Super-resolving Malus' law for the linear photonic gear.}
 \label{fig1}
\end{figure*}

\section*{Results}

Let us consider the situation depicted in Fig.\ \ref{fig1} where we are interested in measuring the relative transverse displacement $\Delta x$,  along the $x$ axis, between two objects (GP1 and GP2). 
To do so we prepare a collimated laser beam in the state $\ket{H}$, that is uniformly linearly polarized along the horizontal direction, and let it pass through a $g$-plate \cite{gplate} (lying in the reference frame of GP1). 
A $g$-plate is a patterned liquid crystal slab where the orientation $\alpha$ of the molecular director follows the geometry:
\begin{equation}\label{alpha}
\alpha(x,y) = \frac{\pi}{\Lambda} x+ \alpha_0 
\end{equation}
where $\Lambda$ is the spatial period of the device and $\alpha_0$ is an offset angle (see Fig.\ \ref{fig1}a). 
At the exit of the $g$-plate, the optical polarization takes the following expression (see SM):
\begin{equation}
\ket{H} \rightarrow \ket{\Lambda} = \cos[2\alpha(x)] \ket{H} + \sin[2\alpha(x)] \ket{V}
\end{equation}
The state $\ket{\Lambda}$ represents  a structured light beam where the polarization direction varies linearly along the $x$ axis with a period that corresponds to half the period of the $g$-plate pattern. Let us now assume that the beam passes through a second $g$-plate (lying in the reference frame of GP2), identical to the first one but laterally shifted by an amount $\Delta x$. As detailed in the SM, the output field is then: 
\begin{equation}\label{teta}
\ket{\Lambda} \rightarrow \ket{\theta} = \cos(\Delta\theta) \ket{H} + \sin(\Delta\theta) \ket{V}
\end{equation}
that is  a homogeneous linearly polarized beam, with the polarization direction rotated by 
an angle $\Delta\theta$ with respect to the input state and where: 
\begin{equation}\label{gears}
\Delta\theta = \frac{2 \pi \Delta x}{\Lambda}.
\end{equation}
Equation (\ref{gears}) represents the  {\it linear gears} mapping between the displacement $\Delta x$ and the polarization rotation $\Delta\theta$.
Crucially, this rotation can be amplified by reducing the value of the $g$-plates spatial period $\Lambda$, as when decreasing the radius of mechanical gears.  Let us stress that this holds for any linearly polarized state, and the input $\ket{H}$ was chosen only for the sake of simplicity. 
Moreover, in order to neglect the beam diffraction when deriving equation (\ref{teta}), we are assuming that the distance $D$ between $g$-plates is sufficiently small, that is $D\ll w_0 {\Lambda/\lambda}$, where $w_0$ is the beam radius at the waist position and $\lambda$ is the optical wavelength. As an alternative, for larger values of $D$, a lens system can be used to image the first $g$-plate onto the second one. 

To measure $\Delta x$, one can read out the optical power after a projection over the initial polarization state,
leading to a super-resolving Malus' law (Fig. 1c):
\begin{equation}\label{gear}
P(\Delta x)=P_0\left|\braket{H|\theta}\right|^2=P_0 \cos ^2 \left(\frac{2 \pi \Delta x}{\Lambda}\right)
\end{equation}
where $P_0$ is the power of the input laser beam. An offset term in the argument of the cosine function (not shown in equation \ref{gear}) can be adjusted by rotating either the input polarization or the polarizer orientation. This in turns allows one to accurately shift the Malus' law curve so as to operate in its linear regions, where the maximum sensitivity $S$ is obtained and we get:
\begin{equation}\label{lin}
P_{lin}\simeq P_0\left(\frac{1}{2}\pm \frac{2\pi}{\Lambda}\Delta x\right).
\end{equation}
This corresponds to a sensitivity $S=\left|\frac{dP}{d\Delta x}\right| = 2{\pi}P_0/\Lambda$, that can be increased by reducing $\Lambda$.
\begin{figure*}[t!]
\includegraphics[width=1\textwidth,keepaspectratio=true]{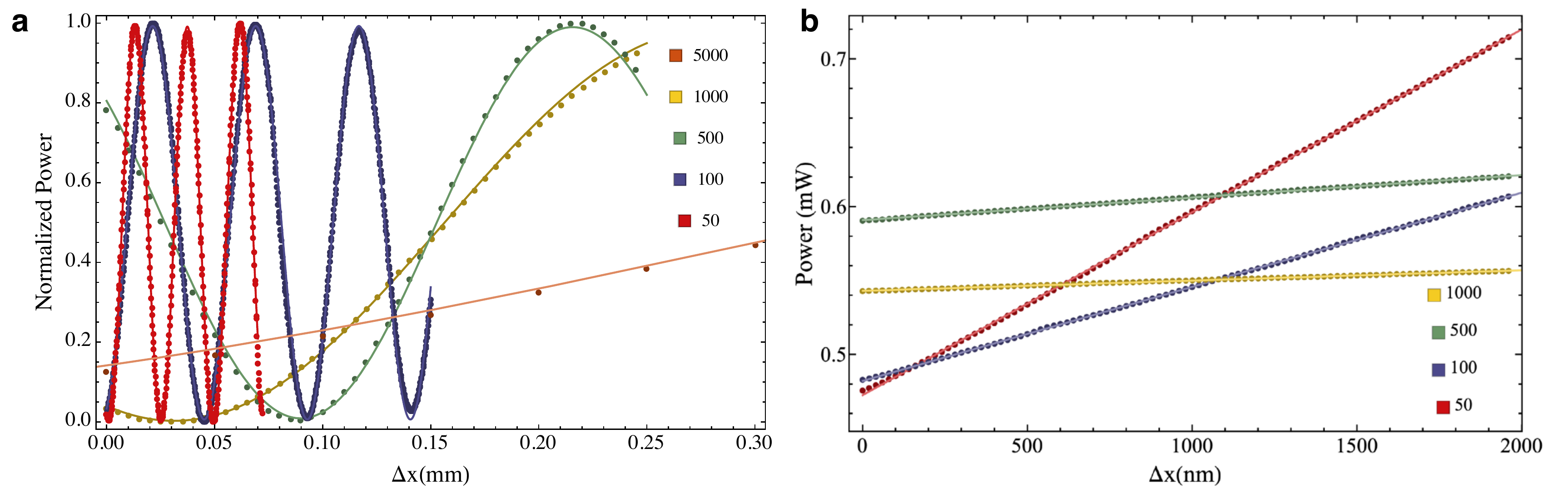}
\caption{Experimental results. a) For each value of $\Lambda$, we reported the normalized measured power of the horizontal polarization component, as a function of the displacement between the two $g$-plates. The zero displacement point for each curve has been set to allow an easy visual comparison between different gears. b) The same measurement is repeated in the linear part of the calibration curves, with a displacement step of $20$ nm. For each configuration, the input optical power was set to a reference value $P_0 = 1$ mW. In both panels, points represent experimental data, solid lines are best fit curves according to equations in the main text, experimental points correspond to the average power over 20 independent measurements (error bars are smaller than the data point size). Specific colors label datasets that are associated with different spatial periods $\Lambda$ (expressed in $\mu$m in the legend).}
 \label{fig3}
\end{figure*}

Importantly, as we increase the sensitivity, we concurrently reduce the working range of the linear gears, which is approximately given by $\Lambda/4$ (monotonicity interval for the Malus' law). For bigger displacements we face, in principle, an ambiguity for the estimation of the correct TD.
However,  
this limit can be overcome in (at least) the following two ways. First, the working point can be always kept in the linear range of the gear by dynamically rotating the polarization analyzer or, equivalently, the input polarization. The degeneracy is then removed by keeping track of this rotation. Second, it is possible to exploit an additional pair of $g$-plates with a period $\Lambda'$ large enough to remove the degeneracy, while the desired resolution is provided by the original $g$-plate pair.


To validate our findings, we realized the setup sketched in Fig.\ \ref{fig1} (more details in SM). Specifically, we sent a collimated He:Ne laser beam ($\lambda$ = 633 nm, horizontally polarized) through a pair of $g$-plates, considering in our experiment five types of pairs, with spatial periods $\Lambda = 5000, 1000, 500, 100, 50$ $\mu$m, respectively. At the exit of the second plate, a polarizer was used to filter the desired linear polarization component and the optical power was measured by a power-meter. To displace the plates in a controllable manner, the second cell was mounted on a motorized translation stage. 
Normalized measured powers for different TDs are reported in Fig.\ \ref{fig3}a, together with the best fit curves according to equation (\ref{gear}). Obtained data nicely reproduce the expected oscillatory behavior.

\begin{figure}[ht]
\centering
\includegraphics[width=0.5\textwidth,keepaspectratio=true]{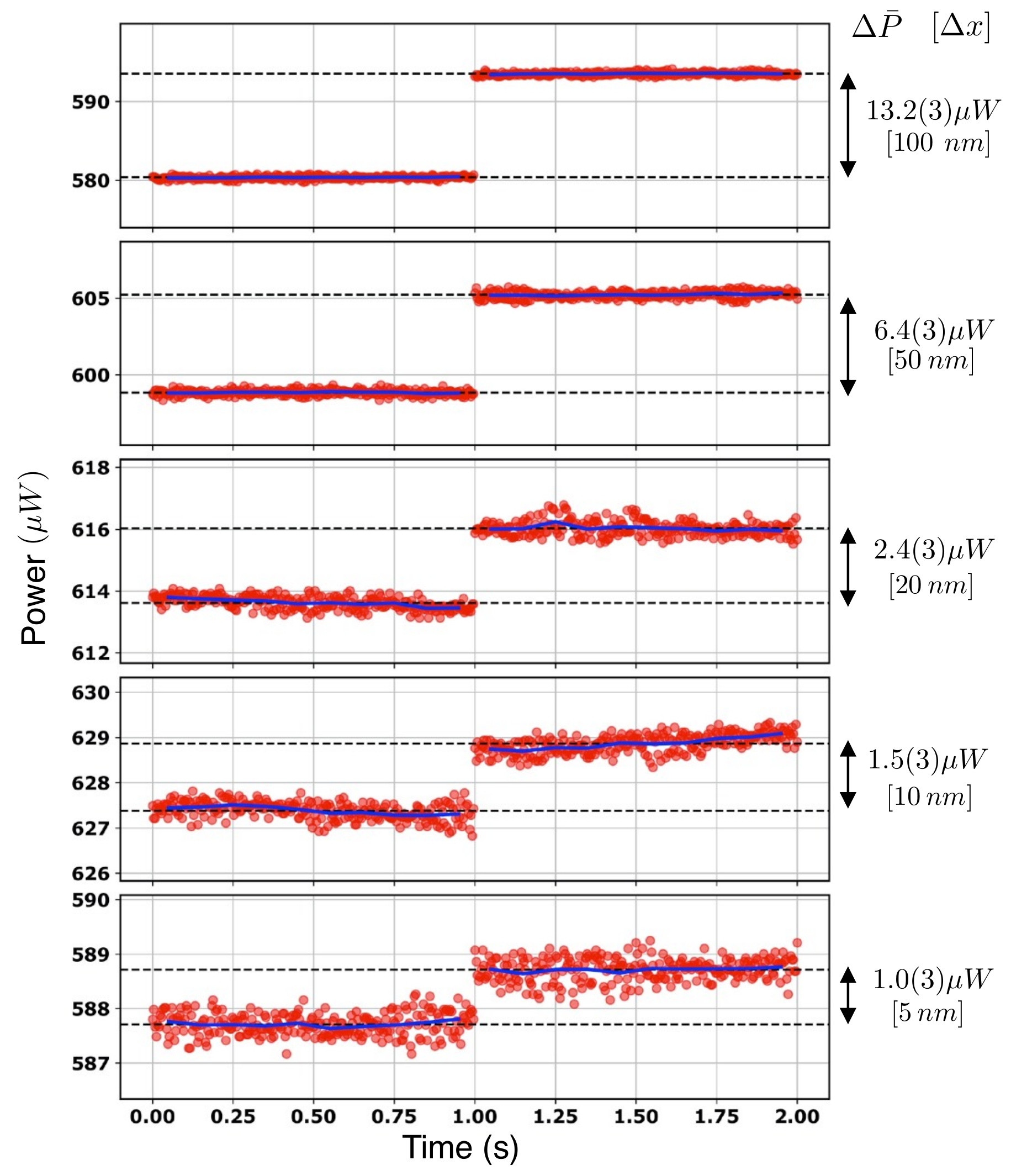}
\caption{Step measurements for gear $\Lambda = 50$ $\mu$m. Each plot reports the optical power measured for 1 second (250 red points) before and after a controlled step $\Delta x$ of the translation stage. Blue lines represent the average power calculated over time intervals of 0.1 $s$ while black dashed lines mark the total average power calculated before and after each step. The corresponding power difference $\Delta \bar{P}$ is reported, together with the step amplitude, on the right side of each plot.}
 \label{fig4}
\end{figure}

To evaluate and compare the sensitivity of different gears, we then focused our measurements on the linear region. We therefore set the optical power to a reference value $P_0 = 1$ mW and measured the calibration curve $P(\Delta x)$ in the linear region, for a total displacement of 2 $\mu$m with steps of 20 nm (see Fig.\ \ref{fig3}b). To improve the plot clearness, each dataset has been measured in a slightly shifted region with respect to the middle of the fringe (but still abundantly in the linear region of the calibration curve).  A comparison between the slopes of the curves clearly shows the advantage in terms of sensitivity that is obtained by decreasing $\Lambda$. This can be readily quantified by performing a linear fit for each curve. The best sensitivity $S = 124.0 \pm 0.1$ nW/nm   
was obtained, as expected, for $g$-plates with $\Lambda = 50$ $\mu$m.  
\
We therefore evaluated the actual resolution of our detection system for this configuration. To this end we repeatedly measured the optical power for a time interval of 1 s (250 points), before and after a controlled step of the translation stage. We report these results in Fig.\ \ref{fig4}, where circles represent experimental data, while black dashed lines mark the average power as calculated over  each set of 250 points. We started with a "large" displacement $\Delta x = 100$ nm, that lead to a mean power difference $\Delta \bar{P} =13.2$ $\mu$W, and  gradually decreased the step size. The smallest measured displacement $\Delta x = 5$ nm, corresponding to approximatively ${\lambda/125}$, is still clearly resolved as the measured $\Delta \bar{P}=1.0$ $\mu$W is significantly larger than its error bar ($\sigma$ =0.3 $\mu$W), calculated as the quadrature sum of the standard deviations of the power distribution before and after the step, respectively.  Further reducing the displacement would not lead to meaningful results, as the nominal displacement accuracy of our translation stage is approximately $2$ nm.    
Let us notice that the power standard deviation when using $g$-plates with $\Lambda =$ 50 $\mu$m was typically $\sigma_P =$ 0.2 $\mu$W, yet it reduced to $\sigma_{Po} =$ 0.1 $\mu$W when the plates were switched off. Therefore, we ascribe the residual value of the standard deviation of our signal to vibrations of the optomechanical components.

An estimate of the system resolution is obtained in terms of the ratio $R=\sigma/S$.
To evaluate this parameter, it is crucial to ged rid of fast signal oscillations that are mainly due to mechanical fluctuations of our system. To this end, we considered the average power over time intervals of 0.1 s (blue lines in Fig. 3), yielding a typical standard deviation $\sigma_P =$ $0.05$ $\mu$W. This corresponds to a sub-nanometric resolution $R \simeq 400$ pm $\left(\lambda/1580\right)$.\\

In case of an ideal target without vibrations, this result could be further improved if fluctuations in optical power  are minimized, as for instance by employing an ultrastable laser or a balanced detector. The value of $\Lambda$, on the other hand, can be decreased down to few microns (or even below if replacing $g$-plates with dielectric metasurfaces \cite{Lin2014}). 
By referring to state-of-the-art liquid crystal technology, we can consider $g$-plates with $\Lambda=6$ $\mu$m  \cite{Koma07} which, if considering the power fluctuations of our system (at $P_0 = 1$ mW) would yield the remarkable resolution of $R\simeq50$ pm, corresponding to approximatively  $\lambda/12500$.

\section*{Discussion}
In this paper we have reported a novel optical encoding technique that enables ultra-sensitive measurements of transverse displacements. This is done by directly mapping displacements between two objects into polarization rotations of a collimated laser beam.
Let us note that, in principle, it is possible to encode TD in modulations of the field amplitude \cite{Hane85} (rather than polarization), by using a pair of amplitude masks instead of $g$-plates. However, as detailed in SM, linear gears outperform amplitude encoding both in terms of sensitivity (by almost an order of magnitude) and stability.
By operating in ambient conditions and room temperature with a He-Ne laser, low optical power and a standard power meter, we demonstrated a resolution of $R = 400$ $pm$, which could plausibly be brought down to $R = 50$ pm by considering existing state-of-the-art liquid crystal fabrication technology.  
Let us remark that this result is obtained with a practical setup that relies on a single optical path, thus not suffering from the typical instabilities that affect metrological interferometric instruments. Our setup could be eventually implemented in a compact sensor with cost effective integrated laser diodes and photodetectors. We envisage the use of this scheme, for both research and industry, in a number of future sensing devices, for instance for monitoring deformations and displacement of precision components or structures, measuring material properties such as elasticity modulus, in nanofabrication or microscopy.

\subsection*{Acknowledgements:}
This work was supported by the Italian Ministry of Education, University, and Research through the PRIN Project INPhoPOL and the European Union Horizon 2020 program, within European Research Council Grant No. 694683, PHOSPhOR. A patent application related to this work is currently pending. 

\onecolumngrid

\clearpage
\renewcommand{\theequation}{S\arabic{equation}}
\renewcommand{\thesection}{S\arabic{section}}
\renewcommand{\thefigure}{S\arabic{figure}}
\renewcommand{\theHfigure}{Supplement.\thefigure}
\renewcommand{\figurename}{Supplementary Figure}

\setcounter{equation}{0}
\setcounter{section}{0}
\setcounter{subsection}{0}
\setcounter{figure}{0}
\setcounter{page}{1}
\begin{center}
\textbf{Supplementary Material}
\end{center}
\vspace{1 EM}

\section{Fabrication of $g$-Plates}
$g$-plates are fabricated using a well established procedure detailed
in the following. Two ITO (indium tin oxide) coated glass slabs ($\mathrm{25 \times
25mm^2}$) are spin coated with a solution of Brillant Yellow, a commercial azo
dye in N,N-Dimethylformamide. Spacer beads with diameter $\mathrm{6{\mu}m}$
are deposited on one of the substrates (ITO side), then the two glass slabs
are glued together (ITO sides inward) to stably maintain the
$\mathrm{6{\mu}m}$ spacing gap. The desired anchoring pattern is written onto the
photo-alignment layer via 1D direct laser writing. The liquid crystal cell is
then filled with a nematic liquid crystal E7. Finally, electric contact are
added (ITO side) in order to provide control on the effective extraordinary
refractive index. At moderate bias voltages, the $g$-plate behave effectively
like an electrically tunable polarization grating.

\section{Working principle details}
The action of a $g$-plate is determined by its birefringent optical retardation $\delta$, whose value can be adjusted by tuning an external alternating voltage \cite{Piccirillo2010}.
By setting $\delta=\pi$, the $g$-plate can be described as a geometric-phase grating whose action on the polarization state can be easily written in the circular polarization basis as:
\begin{equation}\label{gplate}
\ket{L/R}\rightarrow e^{\pm i 2\alpha(x)}\ket{R/L}
\end{equation}
where $\ket{L}$ and $\ket{R}$ stand for left and right circular polarization states respectively and $\alpha(x)$ is a linear function of the transverse position given by Equation (1) of the main text. 
Let us now consider a Gaussian beam $\ket H$, uniformly polarized along the horizontal direction ($x$ axis), and propagating along the $z$ axis. 
The $g$-plate transformation on the input beam results in: 
\begin{equation}
\ket{H} \rightarrow \ket{\Lambda} = \frac{1}{\sqrt{2}} \left(e^{i2\left({\pi}x/\Lambda+\alpha_0\right)} \ket{R} + e^{-i2\left({\pi}x/\Lambda+\alpha_0\right)} \ket{L}\right)
\end{equation}
where $\ket{L/R}$ denote left and right circular polarization states, respectively. Essentially, each circular polarization component is a Gaussian beam, propagating along a direction in the $xz$ plane forming with the $z$ axis an angle that is approximately given by $\simeq\pm \lambda/\Lambda$. As such, in the near field the beam keeps its Gaussian envelope, yet it features a spatially-inhomogeneous polarization pattern (see Fig.\ 1 in the main text). After passing through a second $g$-plate, identical to the first one but laterally displaced by an amount $\Delta x$, the optical field is described by a state:
\begin{equation}\label{tetaSM}
\ket{H} \rightarrow \ket{\theta} = \frac{1}{\sqrt{2}}\left(e^{i \Delta\theta}\ket{R} + e^{-i \Delta\theta}\ket{L}\right) = \cos(\Delta\theta) \ket{H} + \sin(\Delta\theta) \ket{V}
\end{equation}
with $\Delta\theta = 2 \pi\Delta{x}/\Lambda$

\section{Experimental details}

\begin{figure*}[ht]
\centering
\includegraphics[width=0.95\textwidth,keepaspectratio=true]{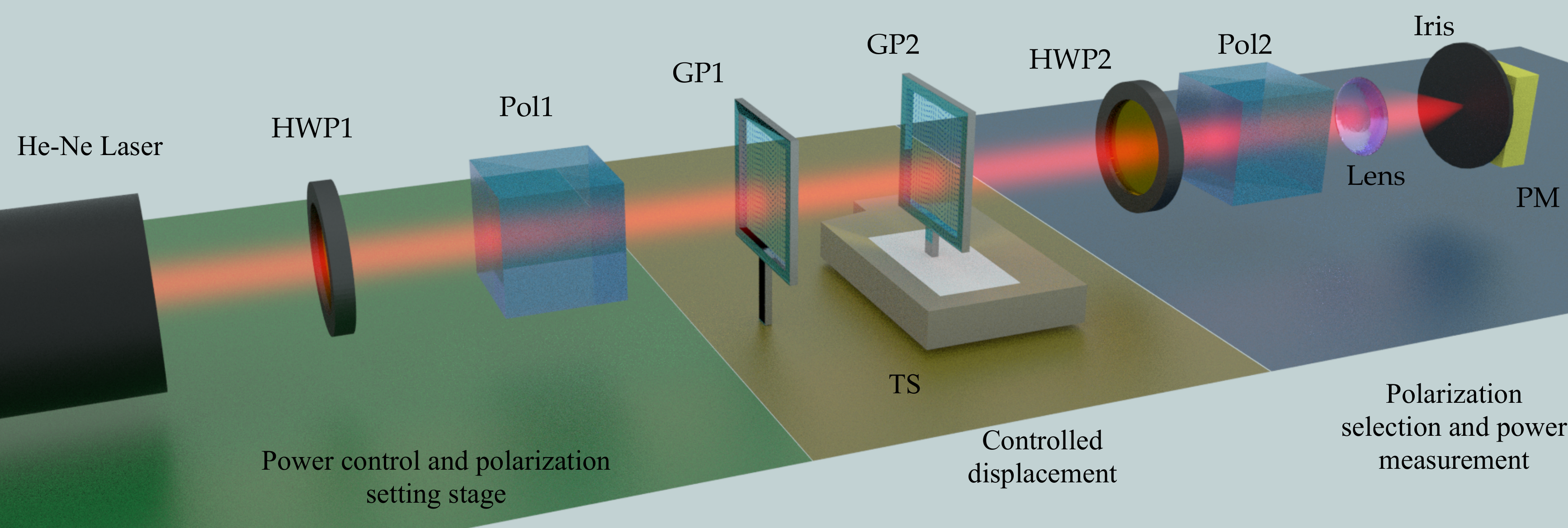}
\caption{Detailed experimental setup: A collimated He-Ne laser beam, is controlled in intensity and set into horizontal polarization state via  a half waveplate (HWP1) and a polarizer (Pol1). The linear gear is implemented with the two $g$-plates (GP1 and GP2). The displacement between the two devices is controlled via a motorized translation stage (TS). A second pair of half waweplate (HWP2) and polarizer (Pol2) allows one to project the polarization along an arbitrary linear state. By rotating HWP2 it is possible to set the working point.  The optical power is measured via a power meter (PM) after a spatial filter (lens + iris) that blocks the unconverted light thus increasing the overall fringe visibility.}
 \label{fig2}
\end{figure*}

To validate our predictions we implemented the linear gear with the experimental setup depicted in Figure (\ref{fig2}). A collimated He-Ne laser beam is initialized in the $\ket{H}$ state by a linear polarizer (Pol1) while a half wave plate (HWP1), placed before Pol1, is used to control the optical power $P_0$.  The linear gear is then implemented via two identical $g$-plates (GP1 and GP2). 
In order to control the transverse displacement $\Delta x$ between the two devices, GP2 is mounted on a motorized translation stage.  Depending on the maximum required TD, we used two different translation stages, one with a position accuracy of $100$ nm and long travel range, and a second one, based on a piezoelectric positioner, with a displacement accuracy of $2$ nm. The power $P(\Delta x)$ is recorded by a power meter (PM) placed after a second polarizer (Pol2) and a spatial filter (Lens + Iris in the focal plane). The latter is used to improve the Malus' law visibility, as it filters out unwanted light components associated with high spatial frequencies (possibly due to inaccuracies in the tuning of $g$-plates or in their pattern). A second half waveplate (HWP2) is placed between Pol2 and GP2 to rotate the analysed polarization direction, in order to set the working point in the desired position.

Finally, let us notice that in order to properly take into account experimental imperfections (mainly polarization purity, $g$-plate conversion efficiency and background noise) equations 5 and 6 need to be slightly modified by adding an extra offset term $P_\mathrm{off}$. This term has no direct effect on the sensitivity as we consider $P_0 = P_\mathrm{tot} - P_\mathrm{off}$, where $P_\mathrm{tot}$ is the total power of the beam reaching the power meter, therefore we omitted it in the main text for the sake of clarity.

\section{Performance}

Here, we derive a comparative performance analysis of linear optical encoders demonstrating the advantage of the linear photonic gears with respect to encoders based on amplitude gratings. For the sake of simplicity, we limit our discussion to sine and binary amplitude gratings, by considering simple
models, able to give the contour for both quantitative and qualitative results, without loss of generality. 

We assume an input Gaussian beam with
power $P_0$, waist  $\omega$, centered at $x_0$, i.e.
$I(x)=P_{0}\sqrt{\frac{2}{\pi\omega^{2}}}e^{-2(x-x_{0})^{2}/\omega^{2}}$. As in the main section, we assume that the two gratings are close enough in order to neglect diffraction effects.
A single $g$-plate is effectively an electrically controlled birefringent plate with varying optical axis. As such, it can
be modeled using Jones matrices (written in left/right handed reference). Here, \[J(\delta, \theta)=\left(
\matrix{\cos\frac{\delta}{2} & i\sin\frac{\delta}{2}~e^{-2i\theta} \cr i\sin\frac{\delta}{2}~e^{2i\theta} &
\cos\frac{\delta}{2}}\right),\] with $\theta$ the orientation of the optical axis, and $\delta$ the effective
phase shift between slow and fast axis. 
For half-wave tuned $g$-plates, $\delta=\pi$, the Jones matrix of the stacked $g$-plated pairs reads as: \[T(\theta_2,
\theta_1) = J(\pi, \theta_2) \cdot J(\pi, \theta_1) = -\left(\matrix{e^{2i(\theta_1-\theta_2)} & 0 \cr 0 &
e^{-2i(\theta_1-\theta_2)}}\right).\] When the second $g$-plate is shifted about $\Delta x$, haven written
$\theta_{2}(x) =\theta(x -x_2-\Delta x)$, $\theta_1(x)=\theta(x-x_1)$, where $x_1$ and $x_2$ represent offsets, with
$\theta(x) = \pi x/\Lambda$, the effective transmission coefficient of a linearly polarized beam, horizontally (without
loss of generality), analyzed through a polarizer at angle $p$ read as: $T(\Delta x)= \cos^{2}(2\pi\Delta x/\Lambda +
2\pi(x_2-x_1)/\Lambda+p)$. The analyzer can be therefore used to  shift the operating point. We then
set $T(\Delta x)= \cos^{2}(2\pi\Delta x/\Lambda)$. Interestingly the effective transmission coefficient is constant
across the exit plane thus the recorded power is simply 
\begin{equation}
P(\Delta x)/P_0 = \cos^{2}(2\pi\Delta x/\Lambda),
\end{equation}
independently from the center and waist, generally from amplitude profile of the input beam.

For sine grating pairs with transmission coefficient $T(x)= \left(1+
\cos\left(2{\pi}x/\Lambda\right)\right)/2$, the recorded intensity for a lateral shift $\Delta x$ is given by:

\begin{equation}\label{eq:Transfer:Sine}
\matrix{
P(\Delta x)/P_{0} & = \frac{1}{4} + \frac{1}{8}\cos\left(\frac{2\pi}{\Lambda}\Delta x\right) + \frac{1}{4}\left[\cos\left(\frac{2\pi}{\Lambda}x_{0}\right) +\cos\left(\frac{2\pi}{\Lambda}(\Delta x -x_{0})\right)\right]e^{-\frac{1}{2}(\pi\omega/\Lambda)^{2}} \cr \cr
& + \frac{1}{8}\cos\left(\frac{2\pi}{\Lambda}(\Delta x -2x_{0})\right)e^{-2(\pi\omega/\Lambda)^{2}}.
}
\end{equation}

For waists larger than the grating period, 
\begin{equation}\label{eq:Transfer:Sine:ideal}
P(\Delta x)/P_{0}=\frac{1}{4} + \frac{1}{8}\cos\left(2\pi\Delta{x}/\Lambda\right).
\end{equation}

For binary gratings pairs with duty cycle $1/2$ (maximum range and sensitivity) centered in $0$, the recorded output power for lateral displacement $|\Delta x| \leq \Lambda/2$ reads as:
\begin{equation}\label{eq:Transfer:Binary}
P(\Delta x)/P_0 =
\cases{\frac{1}{2}\sum\limits_{n=-\infty}^{\infty}\left[ \mathrm{erf}\left(\frac{\sqrt{2}}{\omega}(\Lambda/4+n\Lambda-x_{0})\right) -\mathrm{erf}\left(\frac{\sqrt{2}}{\omega}(-\Lambda/4+n\Lambda-x_{0}+\Delta x)\right) \right], \cr 0\leq\Delta x \leq \Lambda/2\cr\cr
\frac{1}{2}\sum\limits_{n=-\infty}^{\infty}\left[ \mathrm{erf}\left(\frac{\sqrt{2}}{\omega}(\Lambda/4+n\Lambda-x_{0}+\Delta x)\right) -\mathrm{erf}\left(\frac{\sqrt{2}}{\omega}(-\Lambda/4+n\Lambda-x_{0})\right) \right], \cr-\Lambda/2 \leq\Delta x \leq 0.\cr}
\end{equation}

For waists larger than the grating spacing, we have: 
\begin{equation}\label{eq:Transfer:Binary:ideal}
P(\Delta x)/P_0 = 1/2- |\Delta x|/\Lambda.
\end{equation}
From the results above, one can easily grasp the advantages of $g$-plates over these amplitude masks.
In the optimal case, when the waist is larger than the grating period, the linear sensitivity of the sine encoder is
$S_{\mathrm{sine}} = P_{0}\pi/(4\Lambda)$, for the binary $S_{\mathrm{binary}} = P_{0}/\Lambda$. Therefore the linear gears offers a 
factor $8$ and $2\pi$ of improvement in sensitivity over sine and binary gratings, respectively.  

Moreover, for  smaller  values of  the beam  waist ($\omega<\Lambda/2$),  the  recorded intensity depends on the  beam position for the case of amplitude gratings. As a consequence, factors such pointing instabilities or intensity distribution fluctuations in the beam profile, even for typical dimensions smaller than the grating period, could be particularly detrimental in the case of amplitude grating encoders. Intensity distribution fluctuation are particularly relevant in environments (i.e. high precision machining) where dust particles might scatter the laser beam causing position readout errors. Linear photonic gears, on the other hand, are immune to these effects since the recorded intensity (in the case of perfectly tuned g-plates) neither depend on  the laser beam width nor its position.

\end{document}